# bcd mRNA active transport: 3D agent-based modeling

*Marat A. Sabirov, Victoria Yu. Samuta, Alexander V. Spirov*

*Sechenov Institute of Evolutionary Physiology and Biochemistry of the Russian Academy of Sciences (IEPhB RAS), St-Petersburg, Russia*

## Graphical Abstract

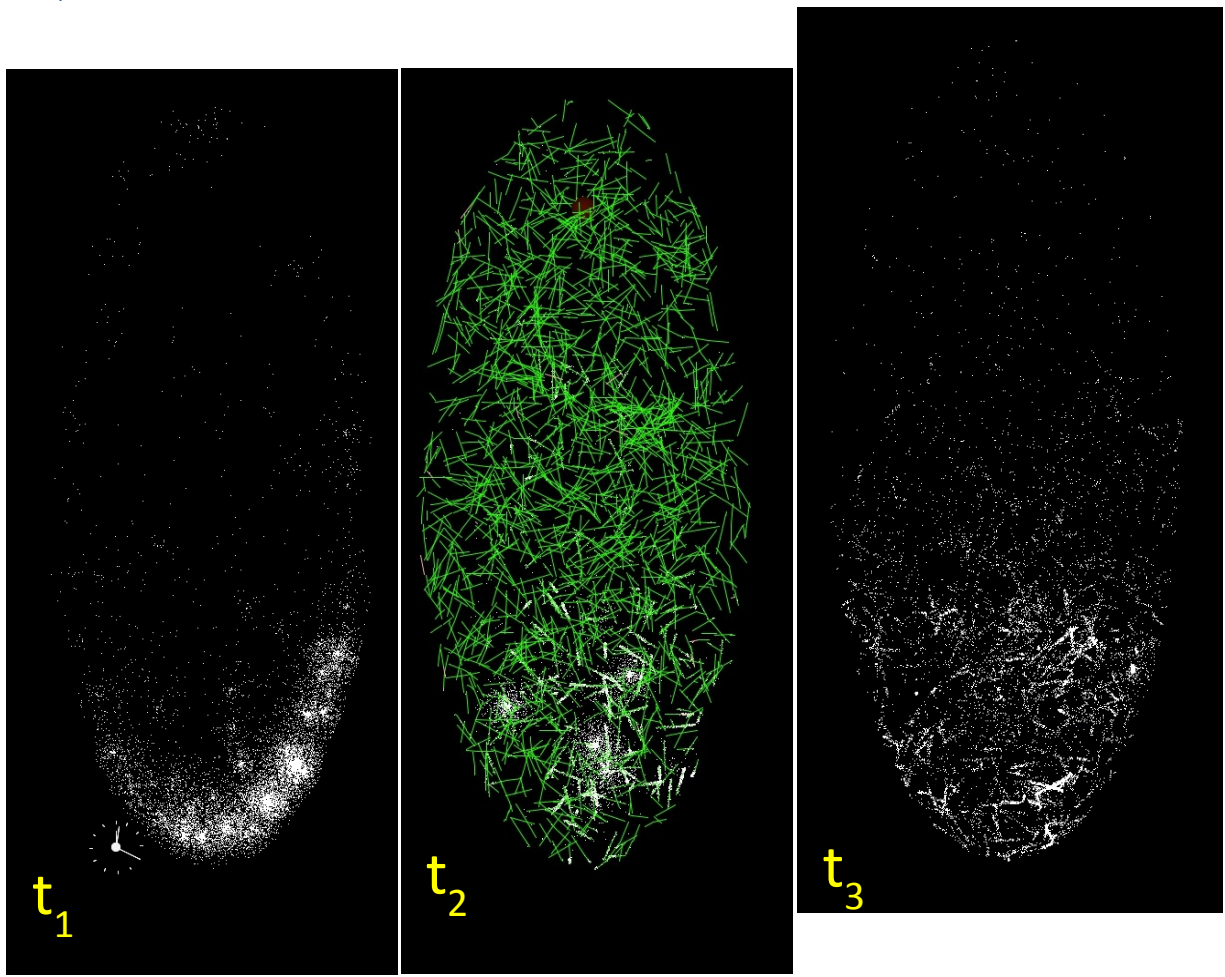

**The 3D agent-based model reproduces active transport by molecular motors (+adaptors) over an undirected network of short microtubules (MT) in syncytial Drosophila embryo.**

## Abstract

The article presents a 3D agent-based model to simulate the dynamics of redistribution of bicoid mRNA. The model is developed on the basis of the Skeledyne software [Odell and Foe, 2008]. The model reproduces active transport by molecular motors over an undirected network of short microtubules. The model is intended to simulate the data presented in [Spirov et al., 2009; Little et al., 2011; Fahmy et al., 2014; Ali-Murthy and Kornberg, 2016; Cai, Spirov, Akber et al., 2017].

# Introduction: 3D agent-based modelling for simulating cytoskeletal behavior \ functions

Agent-based models are computational models simulating actions and interactions of autonomous agents (either individual molecules or collective molecular ensembles). The key notion on which the models are based is that multiple agents interact according to simple rules; these rules are easy to encode in a computer program, and interactions are easy to simulate. However, despite their simplicity, these rules and interactions can generate complex emergent behavior [Mogilner and Manhart, 2016[1]].

3D Agent-based cell modeling allows, for example, to explicitly analyze the dynamics of the cytoskeleton and active transport in the cytoplasm [Odell and Foe, 2008[2]; Wordeman et al., 2016[3]; Mogilner and Manhart, 2016] (See Fig.1).

Here we present a 3D agent-based model to simulate the dynamics of redistribution of bicoid mRNA. The model is developed on the basis of the Skeledyne software developed by Garrett Odell and Victoria Foe [Odell and Foe, 2008]. The model reproduces active transport by molecular motors (+adaptors) over an undirected network of short microtubules (MT) in syncytial Drosophila embryo. The model is intended to make crucial computational tests with the data presented in [Spirov et al., 2009[4]; Little et al., 2011[5]; Fahmy et al., 2014[6]; Ali-Murthy and Kornberg, 2016[7]; Cai et al., 2017[8]].

# Approach

The software developed by Garrett Odell and Victoria Foe from the University of Washington Seattle specifically focuses on the processes of active transport in the cell volume [Odell and Foe, 2008]. We use the software under the terms of GNU General Public License.

Skeledyne software is an agent-based modelling to represent cytoplasm [Odell and Foe, 2008] (Fig.1):

- The authors use LaGrangian "cytoplasmic domains" to keep track of the diffusion and convection of soluble factors throughout a cell's cytoplasm in a way designed to finesse the famously difficult fluid dynamics free boundary problem.
- Agent-based modeling is used to represent the cytoplasm as consisting of thousands of nearly close-packed particles of arbitrary size which collide with their neighbors, which fend each other off (i.e. cannot encroach into the volume occupied by other particles), and which exert viscous drag forces on each other as they move relative to each other [Odell and Foe, 2008].
- Each cytoplasmic domain carries a vector of concentrations of soluble proteins with it as it moves, thus modeling convection. These concentrations represent, for examples, mRNAs being translated into proteins, building block monomers or dimers such as GDP-tubulin and GTP-tubulin dimers from which microtubules polymerize, kinesin motors not yet bound to microtubules, etc [Odell and Foe, 2008].

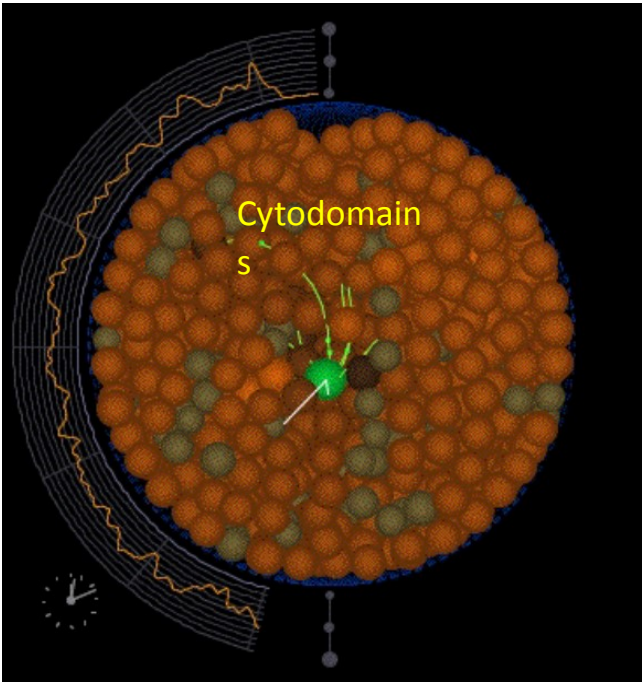

Figure 1. 3D agent-based modelling the early embryo / cells by the Skeledyne software.

## Results and discussion

We model the early Drosophila syncytial embryo as an ellipsoid with axes of 470 * 200 mkm (Fig.2). The surface of the model is mechanically rigid and non-permissive.

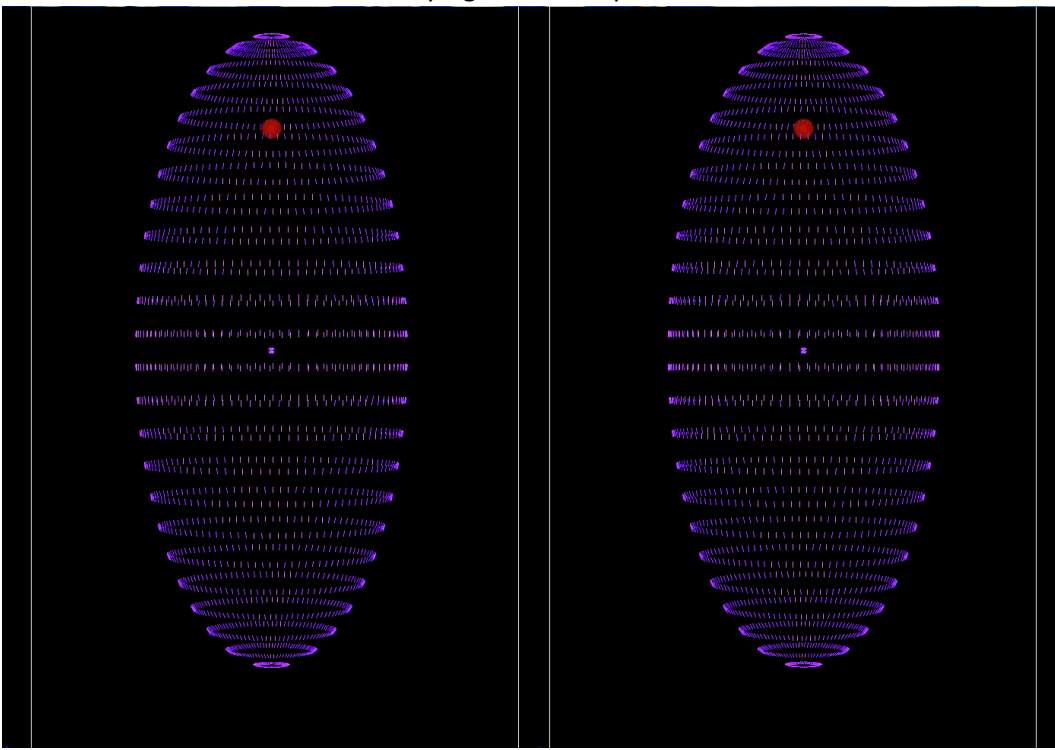

Figure 2. 3D agent-based modeling of the early Drosophila embryo (stereo pair). Cell model is 470 * 200 mkm ellipsoid.

The cargo at the initial moments of the simulation is localized under the surface of the zygote at its head end. The cargo is not secured in any way and will be further redistributed by active transport and free diffusion. The parameters of transport and diffusion are set initially and can be changed on the fly (Fig.3).

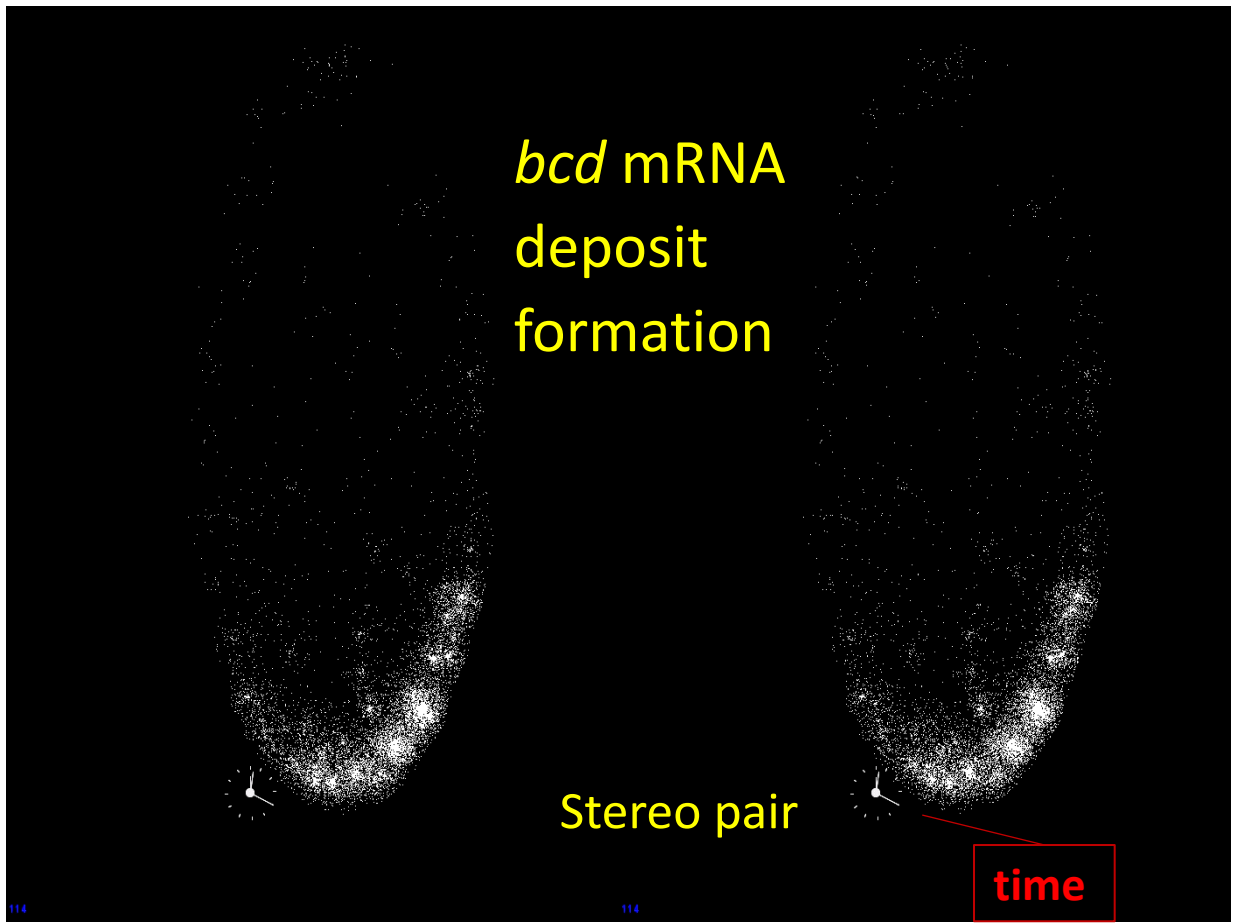

**Figure 3. 3D agent-based modeling of the bcd transport system with 12 thousand cargo complexes (tiny white dots).**

The process of forming an increasingly dense network of MT is arranged as follows. In this version of the model run, the MT assembly all the time exceeds the MT degradation (Fig.4). Therefore, the network over time becomes more and more dense. The MT network occupies the entire volume of the cytoplasm in this version of the model. The length of the MT threads on average is about 40-50 mkm.

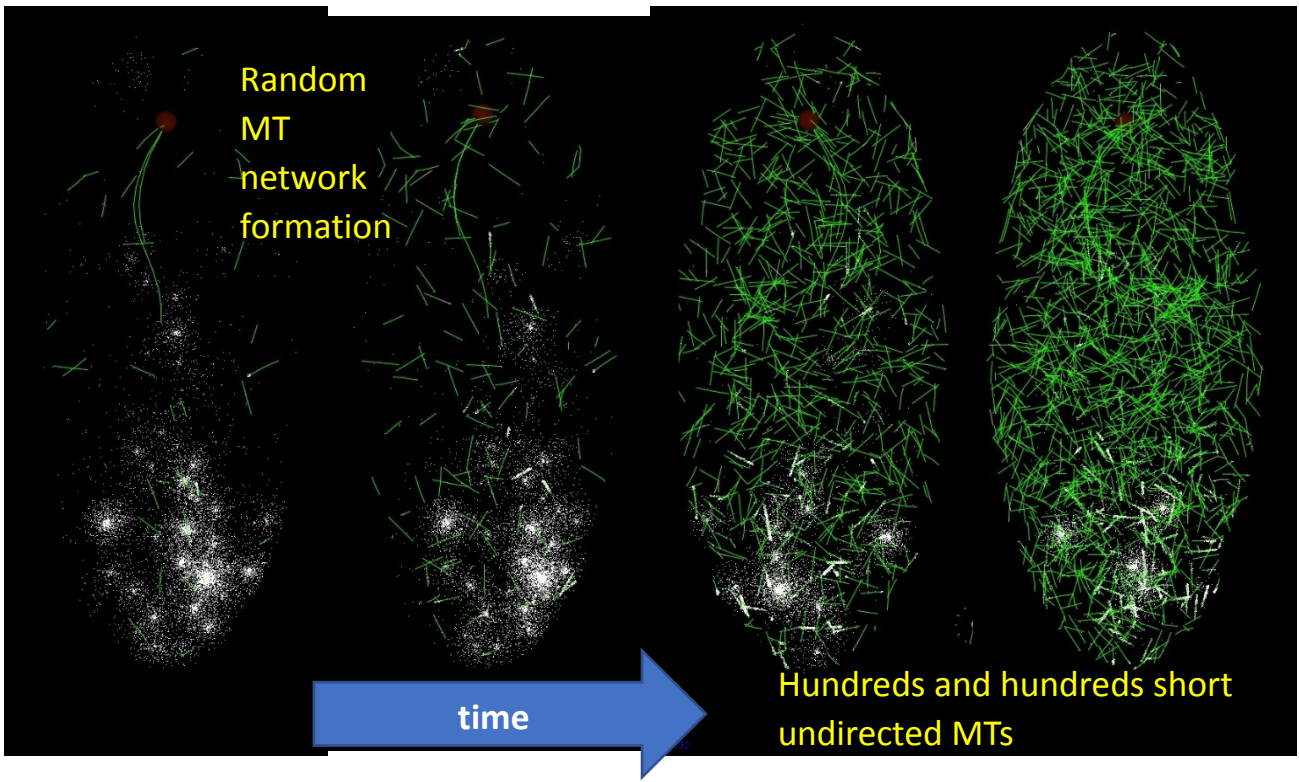

**Figure 4. Modeling of the bcd transport system. Random MT network formation with hundreds short undirected MTs.**

An mRNA molecule, an adapter protein molecule, and a motor form an integral particle. It is modeled as a separate object, capable of jumping onto the MT thread and jumping off of it, actively moving along the MT thread, diffusing passively in the cytoplasm, and so on (Fig.5).

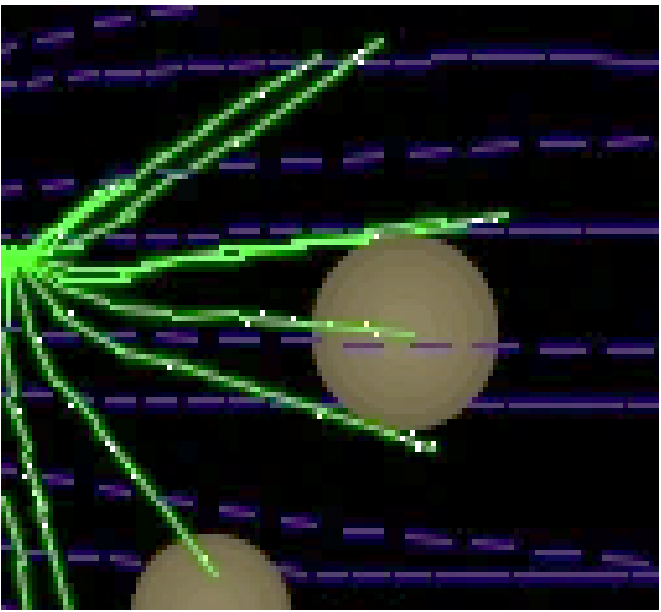

**Figure 5. MT with cargo particles, magnified. MTs are green, the moving cargo particles are white dots.**

Rather quickly, in a matter of minutes, the MT network becomes more and more inhomogeneous: individual MTs stumble into "strands" that are becoming denser and more massive (possibly due to diffusion processes in the cytoplasm of the model) (Fig.6).

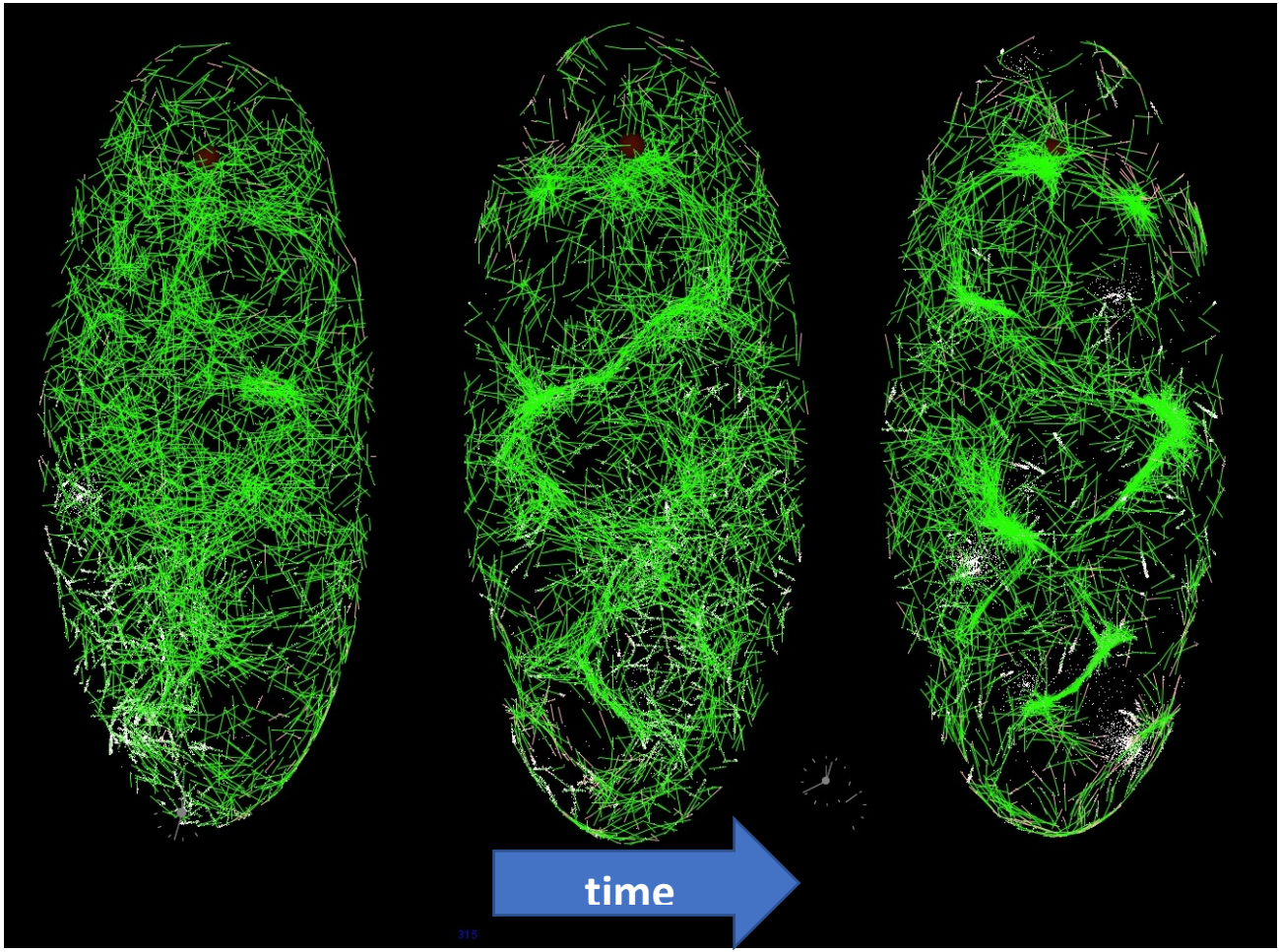

**Figure 6. Random MT network changes: individual MTs stumble into "strands" that are becoming denser and more massive.**

During the model run (approximately 9 minutes of internal model time), the cargo is significantly redistributed from head to tail of the zygote (Fig.7). Herewith, a steep spatial gradient of concentration from head to tail is formed and maintained.

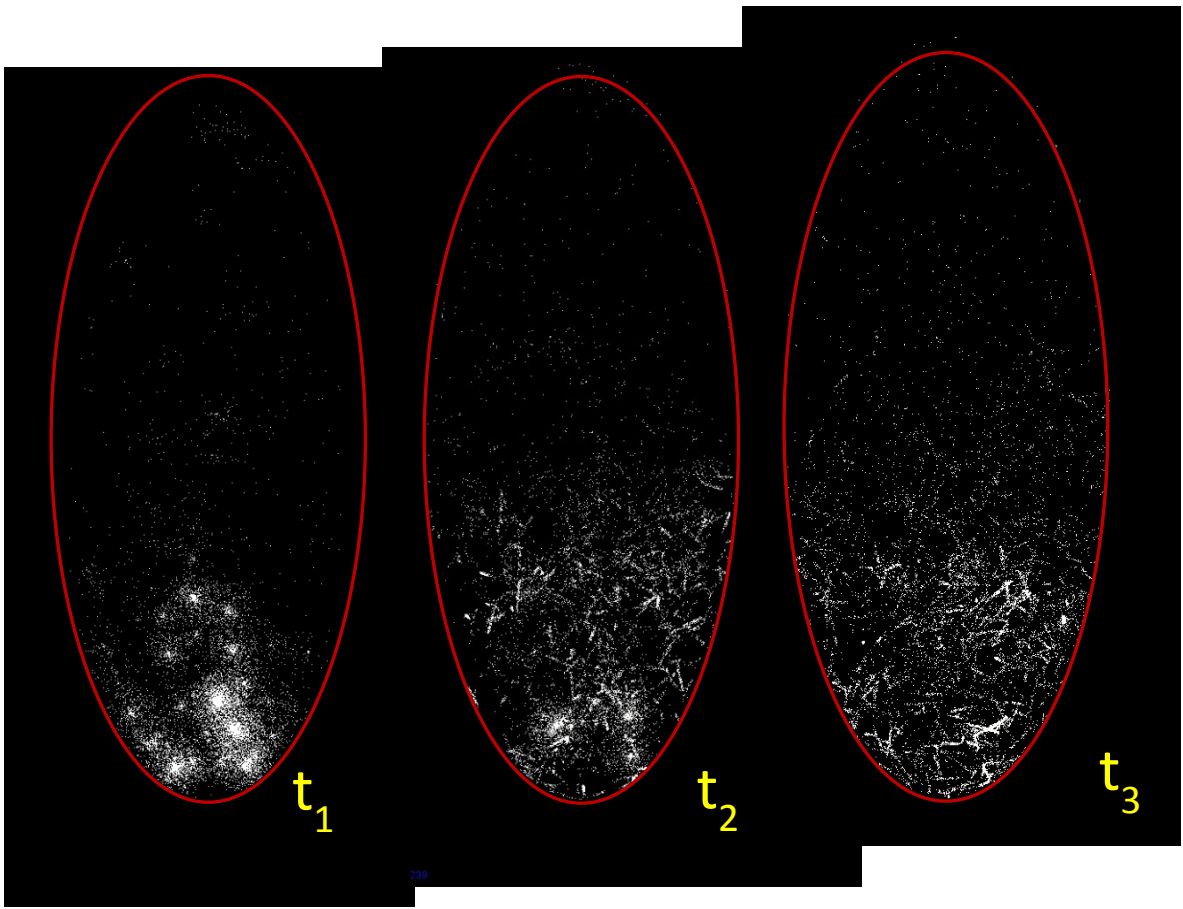

**Figure 7. The cargo transport. A steep spatial gradient of concentration from head to tail is formed.**

Three profiles obtained by processing the three images of the previous figure are shown in Fig.8. Under the chosen modeling conditions, an obvious heterogeneity of distribution and redistribution of the cargo is noticeable.

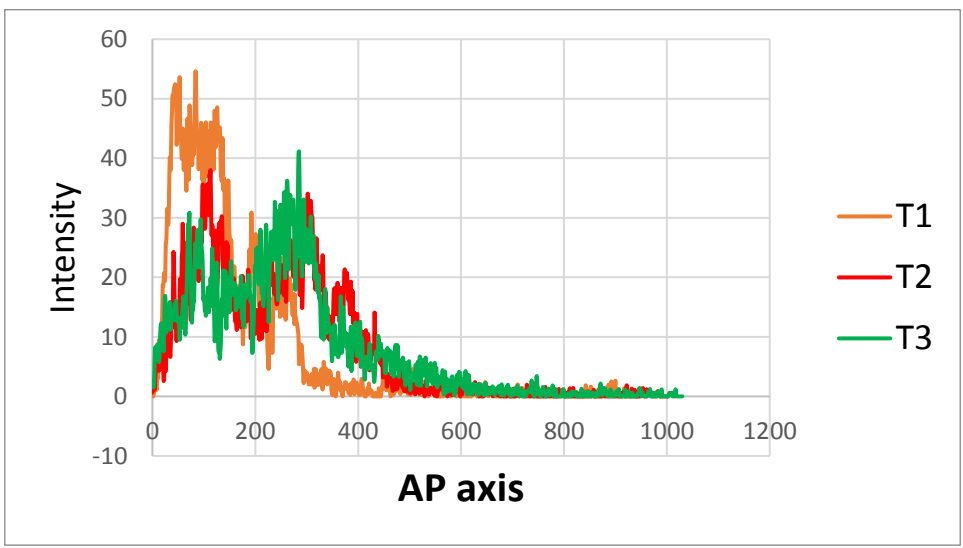

**Figure 8. The cargo transport. The profiles obtained by processing the three images of the Fig.7.**

At last, the ability to add yolk particles will make the bcd transport model more realistic (Fig.9).

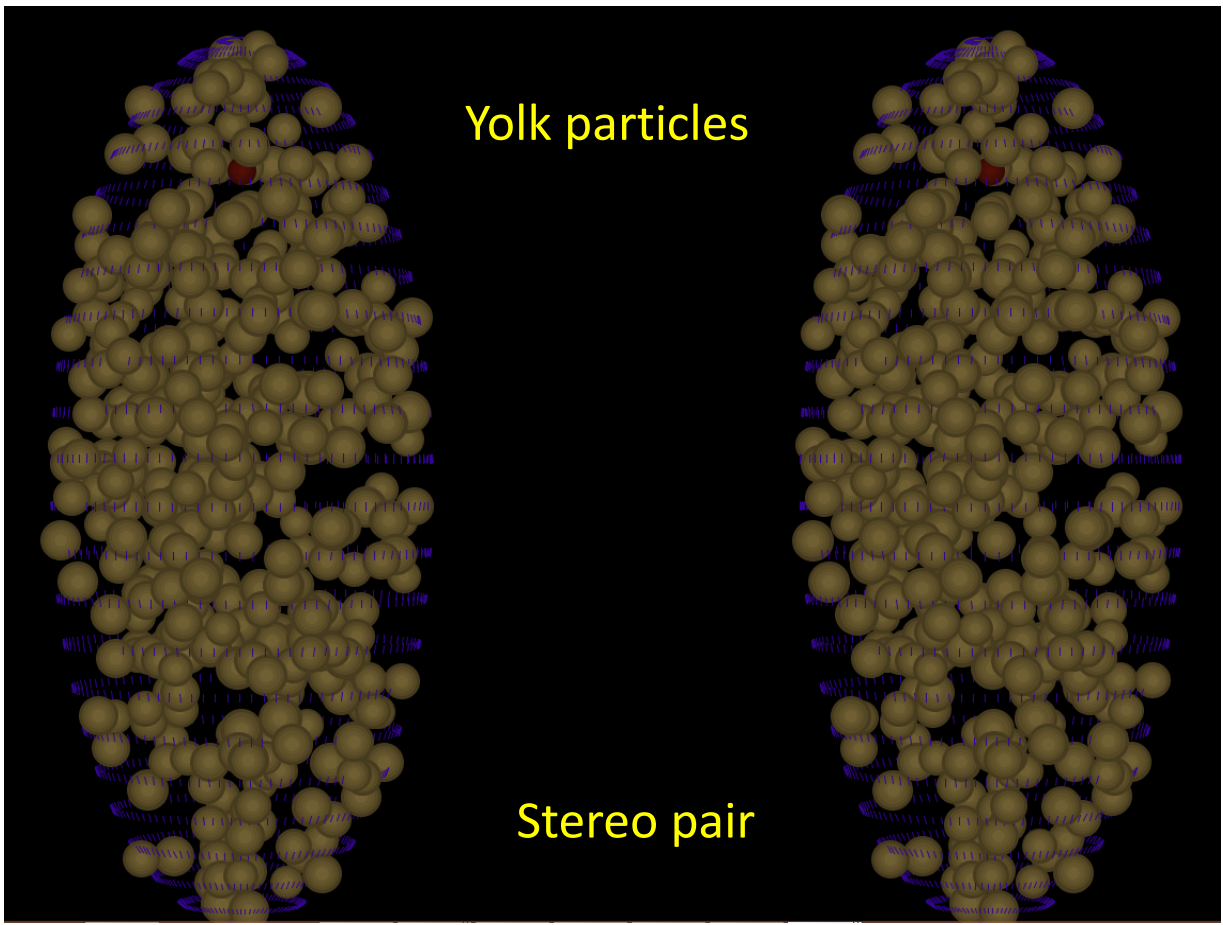

**Figure 9. 3D agent-based modeling with yolk particles.**

## Conclusions

- The behavior of the model in general resembles real biological processes.
- The MT network very quickly (in a matter of minutes of the model time) becomes more and more inhomogeneous: individual MTs stumble into "strands" that are becoming denser and more massive.
- The cargo tends to be redistributed in the embryo volume non-uniformly (possibly due to heterogenization of the MT network).

## Future directions

- We need to figure out how to reduce the heterogenization of the MT network.
- We need to figure out how to achieve a more smooth distribution of cargo.
- It is necessary to reproduce the smooth head-to-tail density gradient MT observed in reality.
- It is necessary to reproduce the processes of active transport in the mass presence of droplets of yolk in the core of the zygote – early embryo.

# References

[1] Mogilner A. and Manhart A. (2016). Agent-based modeling: case study in cleavage furrow models. Mol Biol Cell, 27: 3379-84.

[2] Odell G.M. and Foe V.E. (2008). An agent-based model contrasts opposite effects of dynamic and stable microtubules on cleavage furrow positioning. J Cell Biol, 183(3): 471-483.

[3] Wordeman L., Decarreau J., Vicente J.J., Wagenbach M. (2016). Divergent microtubule assembly rates after short- versus long-term loss of end-modulating kinesins. Mol Biol Cell 27(8):1300-9.

[4] Spirov, A., Fahmy, K., Schneider, M., Noll, M., and Baumgartner, S. (2009). Formation of the bicoid morphogen gradient: an mRNA gradient dictates the protein gradient. DEVELOPMENT 136, 605-614.

[5] Little, S. C., Sinsimer, K. S., Lee, J. J., Wieschaus, E. F., & Gavis, E. R. (2015). Independent and coordinate trafficking of single Drosophila germ plasm mRNAs. Nat Cell Biol. 17(5):558-68.

[6] Fahmy, K., Akber, M., Cai, X., Koul, A., Hayder, A., Baumgartner, S. (2014). αTubulin 67C and Ncd Are Essential for Establishing a Cortical Microtubular Network and Formation of the Bicoid mRNA Gradient in Drosophila. PLoS ONE 9(11): e112053.

[7] Ali-Murthy Z, Kornberg TB. (2016). Bicoid gradient formation and function in the Drosophila pre-syncytial blastoderm. eLife 5 :e13222.

[8] Cai, X., Spirov, A., Akber, M. and Baumgartner, S. (2017). Cortical movement of Bicoid in early Drosophila embryos is actin- and microtubule-dependent and disagrees with the SDD diffusion model, PLOS ONE, 12(10):e0185443.